\documentclass{article}

\newtheorem{theorem}{Theorem}[section]
\newtheorem{prop}[theorem]{Proposition}
\newtheorem{lemma}[theorem]{Lemma}

\def\tr{{\rm tr}}

\def\open#1{\setbox0=\hbox{$#1$}
\baselineskip = 0pt
\vbox{\hbox{\hspace*{0.4 \wd0}\tiny $\circ$}\hbox{$#1$}} 
\baselineskip = 10pt\!}

\def\opens#1{\setbox1=\hbox{${\scriptstyle #1}$}
\baselineskip = 0pt
\vbox{\hbox{\hspace*{0.4 \wd1} $\kern-0.35em {\scriptscriptstyle \circ}$}
\hbox{${\scriptstyle #1}$}} 
\baselineskip = 10pt\!}
 
\def\fn{\open{f}\,}

\def\li{\open{\lambda}\,}
\def\mi{\open{\mu}\,}
\def\lis{\opens{\lambda}}
\def\mis{\opens{\mu}}

\def\dt{\partial_t}
\def\dtheta{\partial_\theta}
\def\dw{ \partial_w}
\def\supp{\mbox{\rm supp}} 
\def\R{{\rm I\kern-.1567em R}}

\begin{document} 
\title{On the Einstein-Vlasov system with hyperbolic symmetry}
\date{}
\author{H\aa kan Andr\'easson\\
        Department of Mathematics\\
        Chalmers University of Technology\\
        S-412\,96 G\"oteborg, Sweden\\
        Gerhard Rein\\
        Institut f\"ur Mathematik, Universit\"at Wien\\
        Strudlhofgasse 4, A-1090 Vienna, Austria\\
        Alan D.~Rendall\\
        Max-Planck-Institut f\"ur Gravitationsphysik\\
        Am M\"uhlenberg 1, D-14476 Golm, Germany  
        }

\maketitle

\begin{abstract}
It is shown that a spacetime with collisionless matter evolving
from data on a compact Cauchy surface with hyperbolic symmetry
can be globally covered by compact hypersurfaces on which the
mean curvature is constant and by compact hypersurfaces on which
the area radius is constant. Results for the related cases of
spherical and plane symmetry are reviewed and extended. The
prospects of using the global time coordinates obtained in this
way to investigate the global geometry of the spacetimes
concerned are discussed.
      
\end{abstract}

\section{Introduction}

One of the key mathematical problems in general relativity is the
determination of the global properties of solutions of the Einstein
equations coupled to various matter fields. We investigate this 
problem in the case of collisionless matter described by the Vlasov
equation. The underlying strategy is to first establish the existence
of a suitable global time coordinate $t$ and then to study the 
asymptotic behaviour of the solution when $t$ tends to its 
limiting values, which might correspond to the approach to a
singularity or to a phase of unending expansion. The following is
almost exclusively concerned with the first of these two steps.
Since the general case is beyond the range of current mathematical 
techniques we study a case with symmetry.

There are several existing results on global time coordinates for 
solutions of the Einstein-Vlasov system. In the spatially homogeneous 
case it is natural to choose a Gaussian time coordinate based on
a homogeneous hypersurface. The maximal range of a Gaussian time
coordinate in a globally hyperbolic solution of the Einstein-Vlasov 
system evolving from data on a compact manifold which are homogeneous 
(or locally homogeneous) was determined in \cite{rendall95b}. It is 
finite for models of Bianchi IX and Kantowski-Sachs types and finite in 
one time direction and infinite in the other for the other Bianchi types. 
All other results presently available on the subject concern spacetimes 
which admit a group of isometries acting on two-dimensional spacelike 
orbits, at least after passing to a covering manifold. The group may be 
two-dimensional (local $U(1)\times U(1)$ symmetry) or three-dimensional 
(spherical, plane or hyperbolic symmetry). In all these cases the 
quotient of spacetime by the symmetry group has the structure of a 
two-dimensional Lorentzian manifold $Q$. The orbits of the group action 
(or appropriate quotients in the case of a local symmetry) are called 
surfaces of symmetry. Thus there is a one-to-one correspondence between 
surfaces of symmetry and points of $Q$.

Three types of time coordinates which have been studied
in the inhomogeneous case are CMC, areal and conformal
coordinates. A CMC time coordinate $t$ is one where each hypersurface 
of constant time has constant mean curvature and on each hypersurface 
of this kind the value of $t$ is the mean curvature of that slice.
In the case of areal coordinates the time coordinate is a function of
the area of the surfaces of symmetry. In some papers in the literature 
it is taken to be proportional to the area. In this paper it is taken to 
be proportional to the square root of the area. In the case of conformal  
coordinates the metric on the quotient manifold $Q$ is conformally
flat. The properties of the last two kinds of time coordinates will
be described in more detail later.

Next the known results concerning these time coordinates in solutions
of the Einstein-Vlasov system will be summarized briefly. For the 
detailed statements the reader is referred to the original papers. 
In the case of spherical symmetry the existence of one compact CMC 
hypersurface implies that the whole spacetime can be covered by a 
CMC time coordinate which takes all real values \cite{rendall95a, 
burnett}. The existence of one compact CMC hypersurface in this case
was proved later by Henkel \cite{henkel01c}, thus providing a complete 
picture in the spherically symmetric case. His proofs use the 
concept of prescribed mean curvature (PMC) foliation which may
be useful more generally. A general local existence theorem for PMC
foliations was proved in \cite{henkel01b}. In the case of 
$U(1)\times U(1)$ symmetry (which includes plane symmetry as a
special case) it was proved in \cite{rendall97a} that the existence of 
one compact CMC hypersurface 
(without loss of generality with negative mean curvature) 
implies that there is a foliation by compact CMC hypersurfaces where the 
mean curvature takes on all negative values. It was also shown that the
foliation covers the whole spacetime between the initial hypersurface and 
the singularity, but it was left open whether it covers the whole spacetime 
in the other time direction. The latter question will be discussed further 
in Section~\ref{results}. In the special case of Gowdy symmetry (which 
includes plane symmetry) the existence of one compact CMC hypersurface 
follows from \cite{henkel01d}. Finally, in the case of hyperbolic symmetry, 
the results of \cite{rendall95a} imply that the existence of one compact 
CMC hypersurface leads to the same conclusions as in the 
$U(1)\times U(1)$-symmetric case, provided the Hawking mass is 
positive on the initial hypersurface. (The question of whether the CMC 
foliation includes all negative numbers as values of the mean curvature
was not discussed in \cite{rendall95a}; it 
will be settled in Section~\ref{results}.) Under the same positivity 
condition the existence of one compact CMC hypersurface is proved in 
\cite{henkel01c}. 

The main concern of this paper is with the areal and conformal
time coordinates. In fact it is the first of these which is of
fundamental interest---conformal coordinates serve as a convenient
tool in an intermediate step in the proofs. Note that the areal time 
coordinate is automatically positive so that the largest possible 
time interval on which the solution could exist is $]0,\infty[$. 
It was shown in \cite{rein94a} for spacetimes with spherical symmetry
that if there is one symmetric Cauchy surface of constant areal time and 
if the data on that hypersurface satisfy a certain inequality then the past 
of the initial hypersurface is covered by an areal time coordinate and  
this coordinate takes on all values in the range $]0,R_1]$, where
$R_1$ is its value on the initial hypersurface. A result improving this
time interval cannot be expected since these spacetimes tend to recollapse, 
and then the areal coordinate will break down at some point in the time
direction corresponding initially to expansion. The results on the
areal coordinate in \cite{rein94a} for the case of plane symmetry
are superseded by those on Gowdy symmetry in \cite{andreasson99a}. 

For Gowdy symmetry the spacetime is globally covered by an areal time 
coordinate which takes all values in the range $]R_0,\infty[$ for some 
$R_0\ge 0$. In some cases, such as the vacuum Gowdy spacetimes, it is 
known that $R_0=0$, cf.\ \cite{moncrief81}, but it is not understood in 
what generality this holds. For hyperbolic symmetry a direct analogue of 
the result for restricted data in spherical symmetry stated above was 
also proved in \cite{rein94a}.    

In the case of hyperbolic symmetry two main questions were
left open by the work up to now. Firstly, is there always one symmetric 
Cauchy surface of constant areal time? Secondly, is there an areal time 
coordinate whose level surfaces are compact and symmetric which
covers the whole spacetime, and does it take on arbitrarily large values? 
These questions will be answered in the affirmative in the following
(Theorem \ref{theorem1}).
The proofs are modelled on the approach of \cite{andreasson99a}, which in
turn was inspired by the work of \cite{berger97} on the vacuum case.
A conformal time coordinate is used in an intermediate step.

In Section~2 the definitions of hyperbolic symmetry and 
the Einstein-Vlasov system are recalled. The local existence of 
conformal coordinates in a neighbourhood of the initial hypersurface 
is demonstrated. In Section~3 the evolution of the spacetime in the 
expanding direction is analysed. A global existence theorem in areal 
time is proved which answers the second main question modulo the first 
one. In the section after that a long-time existence result in the 
contracting direction is proved using conformal coordinates. In Section~5 
the existence of a symmetric compact Cauchy surface of constant area 
radius is demonstrated, thus answering the first main question. All these 
results are combined to produce the main results of the paper in Section~6.
Theorem \ref{theorem1} asserts the existence of a global foliation by 
hypersurfaces of constant areal time in solutions of the Einstein-Vlasov
system with hyperbolic symmetry, and Theorem \ref{theorem2} makes an analogous
statement for hypersurfaces of constant mean curvature. The last section 
discusses possible extensions of these results and looks at what is known 
about the asymptotic behaviour of solutions.

\section{Hyperbolic symmetry}\label{hyperbolic}

In \cite{rendall95a} a definition of spacetimes with surface symmetry 
was given. This comprised three cases, namely spherical, plane and 
hyperbolic symmetry. In the following the case of most interest is 
that of hyperbolic symmetry. The spacetime $M$ is diffeomorphic to 
$\R\times S^1\times F$ where $F$ is a compact orientable surface 
of genus greater than one. The manifold $M$ has a covering space
$\tilde M$ diffeomorphic to $\R\times S^1\times\tilde F$ with projection 
$\hat p$ induced by the projection $p$ from the universal cover $\tilde F$ 
to $F$ according to $\hat p(x,y,z)=(x,y,p(z))$. The surface $F$ admits
metrics of constant negative curvature, and $\tilde F$ endowed with the
pull-back of any one of these metrics is isometric to the hyperbolic plane. 
Let $G$ be the identity component of the isometry group of the hyperbolic
plane. If $\tilde F$ is identified with the hyperbolic plane then an
action $\phi$ of $G$ on $\tilde F$ is obtained. Define an associated
action $\hat\phi$ on $\tilde M$ by $\hat\phi(x,y,z)=(x,y,\phi (z))$.
A spacetime with underlying manifold $M$ defined by a metric 
$g_{\alpha \beta}$ 
and matter fields is said to have hyperbolic symmetry if the pull-back 
$\hat g_{\alpha \beta}$ of the metric and the pull-back of the matter 
fields to $\tilde M$ 
via $\hat p$ are invariant under $\hat\phi$. The surfaces in $M$ 
diffeomorphic to $F$ defined by the product decomposition will be called 
surfaces of symmetry. A Cauchy surface $S$ will be called symmetric if it 
is a union of surfaces of symmetry. Then $\tilde S=p^{-1}(S)$ is invariant 
under the action of the group $G$. It is now clear how to define abstract 
Cauchy data for the Einstein-matter equations with hyperbolic symmetry. They 
should be defined on a manifold $S$ of the form $S^1\times F$ and should be 
such that the pull-back of the data to $\tilde S=S^1\times\tilde F$ under 
the natural covering map is invariant under the natural action of $G$.
The quotient of $\tilde S$ by the action of $G$ is diffeomorphic to $S^1$.

Consider now a choice of matter fields for which the Cauchy problem for
the Einstein-matter equations is well-posed. The example of interest in
the following is that of collisionless matter satisfying the Vlasov 
equation. Corresponding to initial data for the Einstein-matter equations
with hyperbolic symmetry there are data on the covering manifold $\tilde S$ 
which have a maximal Cauchy development on a manifold $\tilde M$. We will
now construct a certain coordinate system on a neighbourhood of the initial
hypersurface in $\tilde M$. As remarked in \cite{rendall95a} coordinates
can be introduced on the initial hypersurface $\tilde S$ so that the
metric takes the form $A^2(\tilde\theta)d\tilde\theta^2
+B^2(\tilde\theta)d\Sigma^2$, where
$d\Sigma^2=dx^2+\sinh^2 x\, dy^2$. Correspondingly there are local 
coordinates on $S$ (local in $x$ and $y$) where the metric takes this form. 
On a neighbourhood of the initial hypersurface in $\tilde M$ we can 
introduce Gauss coordinates based on the given coordinates on $\tilde S$.
The group $G$ acts as a symmetry group on the
initial data. On general grounds the action of $G$ on $\tilde S$ extends 
uniquely to an action on the maximal Cauchy development $\tilde M$ by 
symmetries (see e.g. \cite{friedrich00}). The hypersurfaces of constant 
Gaussian time are invariant under the action of $G$, as are the hypersurfaces 
of constant $\tilde\theta$. Using the isotropy group of any point it can
be seen that the $\tilde\theta x$ and $\tilde\theta y$ 
components of the metric of 
these hypersurfaces must vanish. In the same way, the restriction of the 
second fundamental form of the hypersurfaces to the surfaces of symmetry 
must be proportional to the metric with a factor depending only on $t$ and 
$\tilde\theta$. By integration in $t$ it then follows that 
the spacetime metric takes the form
\begin{equation}
ds^2=-dt^2+A^2(t,\tilde\theta)d\tilde\theta^2+B^2(t,\tilde\theta)d\Sigma^2
\end{equation}
in the region covered by Gauss coordinates. From this we see that we can 
form the quotient of this metric by the action of a discrete group to get 
a spacetime on a subset of $M$ which is a Cauchy development of the original 
data on $S$. Furthermore we can form the quotient by $G$ to get the 
Lorentz metric $-dt^2+A^2(t,\tilde\theta)d\tilde\theta^2$ on a subset 
$Q$ of $\R\times S^1$ referred to in the
introduction. On $Q$ we can pass to double null coordinates $(u,v)$ on a 
neighbourhood of the quotient of the initial hypersurface. Defining new
coordinates by $t=\frac{1}{2}(u-v)$ and $\theta=\frac{1}{2}(u+v)$ puts
the metric on $Q$ in conformally flat form. By pull-back these define
new coordinates on $M$ where the metric takes the form
\begin{equation}\label{cmetric}
ds^2=e^{2\eta}(-dt^2+d\theta^2)+R^2 (dx^2+\sinh^2 x\, dy^2)
\end{equation}
where $\eta$ and $R$ are functions of $t$ and $\theta$ which are periodic
in $\theta$. This proves the existence of a conformal coordinate system 
close to the initial hypersurface. It is possible to choose the double null 
coordinates in such a way that the initial hypersurface coincides with $t=0$ 
and the period of the functions $\eta$ and $R$ is one.

To conclude this section we formulate the Einstein-Vlasov system which governs
the time evolution of a 
self-gravitating collisionless gas in the context of general relativity;
for the moment we do not assume any symmetry of the spacetime.
All the particles in the gas are assumed to have the same
rest mass, normalized to unity, and to move forward in time
so that their number density $f$ is a non-negative function
supported on the mass shell 
\[
PM := \left\{ g_{\alpha \beta} p^\alpha p^\beta = -1,\ p^0 >0 \right\},
\]
a submanifold of the tangent bundle $TM$ of the space-time manifold $M$
with metric $g_{\alpha \beta}$. We use coordinates $(t,x^a)$ with zero 
shift and corresponding canonical momenta $p^\alpha$;
Greek indices always run from 0 to 3, and Latin ones from 1 to 3.
On the mass shell $PM$ the variable $p^0$ becomes a function of the 
remaining variables $(t,x^a,p^b)$:
\[
p^0 = \sqrt{-g^{00}} \sqrt{1+g_{ab}p^a p^b} .
\] 
The Einstein-Vlasov system now reads
\[
\partial_t f + \frac{p^a}{p^0} \partial_{x^a} f - \frac{1}{p^0}
\Gamma^a_{\beta \gamma} p^\beta p^\gamma  \partial_{p^a} f = 0,
\]
\[
G^{\alpha \beta} = 8 \pi T^{\alpha \beta},
\]
\[
T^{\alpha \beta}
= \int p^\alpha p^\beta f \,|g|^{1/2} \,\frac{dp^1 dp^2 dp^3}{-p_0}
\] 
where $\Gamma^\alpha_{\beta \gamma}$ are the Christoffel symbols, 
$|g|$ denotes the determinant of the metric,
$G^{\alpha \beta}$ the Einstein tensor, and $T^{\alpha \beta}$ is
the energy-momentum tensor. The pull-back of the number density $f$ to 
$T\tilde M$ is assumed to be invariant under the action induced on 
$T\tilde M$ by the action of $G$ on $\tilde M$, a fact which can be used 
to reduce the number of independent variables.

\section{The expanding direction}\label{expandd}
\setcounter{equation}{0}

In this section we want to investigate the Einstein-Vlasov
system with hyperbolic symmetry in the expanding direction.
We write the system in areal coordinates, i.e., the coordinates are
chosen such that $R=t$. The circumstances under which coordinates of this
type exist are discussed later. We prove that for initial data on 
a hypersurface of constant time corresponding solutions exist for all 
future time with respect to the areal time coordinate. It should be noted 
that our time coordinate has the geometric meaning of the curvature radius 
of the hyperbolic spaces which form the orbits of the symmetry action. 
Since it requires little additional effort we include for the sake of 
comparison the case of plane symmetry and write the metric in the form
\begin{equation} \label{metric}
ds^2 = - e^{2\mu} dt^2 + 
e^{2\lambda} d\theta^2 + t^2( dx^2 + \sin_\epsilon^2 x\, dy^2),
\end{equation}
where $\mu$ and $\lambda$ are functions of $t$ and $\theta$,
periodic in $\theta$ with period $1$, and
\[
\sin_\epsilon x := \left\{ \begin{array}{ccl}
\sin x&\ \mbox{for}\ &\epsilon = 1,\\
1&\ \mbox{for}\ &\epsilon = 0,\\
\sinh x&\ \mbox{for}\ &\epsilon =-1.        
\end{array} 
\right.
\]
For the case $\epsilon = 1$ the orbits of the symmetry
action are two-dimensional spheres. In this spherically symmetric case
the global result below
is easily seen to be false, cf.\ \cite{rein94a}, so this case
will not be considered further. 
For the case $\epsilon = 0$ the orbits of
the symmetry action are flat tori, and the coordinates
$x$ and $y$ range in the interval $[0, 2\pi]$.

\begin{lemma}\label{lift} 
Assume that (the pull-back of) $f$ is invariant under the group action on 
$T\tilde M$ associated to hyperbolic symmetry. Then $f$ depends only on
\[
t,\ \theta,\ p^1,\ (p^2)^2 + \sinh^2 x\, (p^3)^2.
\]
\end{lemma}

Actually, we will write
\[
f=f(t,\theta,w,L)
\]
where
\[
w := e^\lambda p^1,\ L:= t^4 ((p^2)^2 + \sinh^2 x\, (p^3)^2).
\]
In these variables 
\[
p^0 = e^{-\mu} \sqrt{1+w^2 + L/t^2} =: e^{-\mu} \langle p \rangle ,  
\]
and $L$ is a conserved quantity along particle orbits.
The analogous result holds in the case of plane symmetry $\epsilon =0$.

\noindent
{\em Proof of Lemma~\ref{lift}.}
To see the above assertion on the form of $f$ we proceed as follows.
In the coordinates
\[
\bar x^0=\theta,\ \bar x^1=t \cosh x,\ \bar x^2 = t \sinh x \cos y,\ 
\bar x^3 = t \sinh x \sin y
\]
on the covering manifold $\tilde M$ the symmetry orbits are given as 
$\{\bar x^0 = const\} \times H^2(t)$ where
\[
H^2(t) :=
\{(\bar x^1,\bar x^2,\bar x^3)\in \R^3 \mid \bar x^1>0,\ 
-(\bar x^1)^2 + (\bar x^2)^2 + (\bar x^3)^2 = -t^2 \}, 
\]
a hyperbolic space of radius $t$,\ cf.\ \cite[p.~108 ff]{ON}.
The reason for using these coordinates is that the isometry group
of the hyperbolic spaces now has a nice representation:
Its elements are given by the restriction to $H^2(t)$ of
the linear maps 
$\bar x=(\bar x^1,\bar x^2,\bar x^3)^{\rm tr} \mapsto S \bar x$ 
with matrices $S\in {\rm GL}(3,\R)$
which have the property that
\[
\langle Su,Sv\rangle = \langle u,v\rangle,\ u,v \in \R^3
\]
where
\[
\langle u,v\rangle := -u^1 v^1 + u^2 v^2 + u^3 v^3
\]
is the usual inner product on three-dimensional Minkowski space.
In addition, the matrices $S$ are required to preserve orientation, i.~e.,
\[
S=\left(
\begin{array}{cc}
a_T&b\\
c&a_S
\end{array}
\right)
,\ a_T>0, a_S \in \R^{2\times 2}\ \mbox{with}\ \det a_S > 0,\ b, c \in \R^2.
\]
That these matrices represent the isometries of the hyperbolic spaces
can be found in \cite[p.~239 ff]{ON} or can be seen directly:
In the new coordinates
\[
\bar g_{00}=e^{2\lambda},\ \bar g_{0a}=0,\ \bar g_{ab}=\epsilon_{ab} + 
(1-e^{2\mu})\frac{\delta_{ac}\bar x^c \delta_{bd}\bar x^d}{t^2} \gamma_{ab} 
\]
where
\[
\epsilon_{11}=-1,\ \epsilon_{22} = \epsilon_{33} = 1,\ 
\epsilon_{ab}=0,\ a\neq b,
\]
and
\[
\gamma_{ab}=\gamma_{ba}=-1,\ a=1,b=2,3,\ \gamma_{ab}=1\ \mbox{otherwise},
\]
and since for the matrices introduced above one has
$-S_{1a} S_{1b} + S_{2a} S_{2b} + S_{3a} S_{3b} = \epsilon_{ab}$, 
i.~e., their columns
are orthonormal with respect to $\langle\cdot,\cdot\rangle$, their isometry
property is easily seen. Now denote by $\bar p^\alpha$
the canonical momenta corresponding to the new coordinates
$\bar x^\alpha$, i.~e., 
\begin{equation} \label{vectrans}
\bar p^\alpha = p^0 \frac{\partial \bar x^\alpha}{\partial t} + 
p^1 \frac{\partial \bar x^\alpha}{\partial \theta} + 
p^2 \frac{\partial \bar x^\alpha}{\partial x} +
p^3 \frac{\partial \bar x^\alpha}{\partial y},
\end{equation}
in particular $\bar p^0 = p^1$.
Fix some $\theta=\bar x^0, t>0$, 
$\bar x\in H^2(t)$,
$\bar p^0 \in \R$, and 
$\bar p = (\bar p^1,\bar p^2,\bar p^3)^{\rm tr} \in \R^3$. 
Let $S$ be an isometry such that $S\bar x=(t,0,0)^{\rm tr}$ and
$T$ a rotation in the $(\bar x^2,\bar x^3)$-plane such that 
$TS\bar p=(q^1,q^2,0)^{\rm tr}$
with $q^2>0$. Since $TS$ represents an isometry under which $f$ should be 
invariant,
\[
f(\theta,\bar x,\bar p^0,\bar p)=f(\theta,TS\bar x,p^1,TS\bar p)=
f(\theta,t,0,0,p^1, q^1,q^2,0)
\]
and
\[
q^1 = -\frac{1}{t}\langle TS\bar x,TS\bar p\rangle = 
- \frac{1}{t}\langle \bar x,\bar p\rangle,
\]
\[
(q^2)^2 = (q^1)^2 - (\bar p^1)^2 + (\bar p^2)^2 + (\bar p^3)^2 .
\]
When these quantities are re-expressed in the old momentum variables 
via (\ref{vectrans}) we find
\[
q^1 = p^0,\ (q^2)^2 = t^2 (p^2)^2 + t^2 \sinh^2 x\, (p^3)^2
\]
which proves the assertion in the hyperbolic case. For the plane case 
the argument is even easier.

In the variables which we have now introduced
the complete Einstein-Vlasov system
reads as follows:
\begin{equation} \label{av}
\dt f +\frac{ e^{\mu - \lambda} w}{\langle p \rangle} \dtheta f -
\left( \lambda_t w + e^{\mu - \lambda} 
\mu_\theta \langle p \rangle \right) \dw f =0,
\end{equation}
\begin{eqnarray} 
e^{-2\mu} (2 t \lambda_t + 1) + \epsilon 
&=&
8\pi t^2 \rho , \label{af1}\\
e^{-2\mu} (2 t \mu_t - 1) - \epsilon 
&=& 
8\pi t^2 p, \label{af2} 
\end{eqnarray}
\begin{equation}
\mu_\theta = 
- 4 \pi t e^{\mu + \lambda} j, \label{af3}
\end{equation}
\begin{equation}
e^{- 2 \lambda} \Bigl(\mu_{\theta \theta} + 
\mu_\theta (\mu_\theta - \lambda_\theta)\Bigr)
- e^{-2\mu}\Bigl(\lambda_{tt} + (\lambda_t + 1/t)(\lambda_t - \mu_t)\Bigr) 
=
4 \pi q, \label{af4}
\end{equation}
where
\begin{eqnarray}
\rho(t,\theta) 
&:=& 
\frac{\pi}{t^2} \int_{-\infty}^\infty \int_0^\infty
\langle p \rangle f(t,\theta,w,L)\,dL\,dw = 
e^{-2\mu}T_{00} (t,\theta),\label{ar}\\
p(t,\theta) 
&:=& 
\frac{\pi}{t^2} \int_{-\infty}^\infty \int_0^\infty
\frac{w^2}{\langle p \rangle} f(t,\theta,w,L)\,dL\,dw =
e^{-2\lambda}T_{11} (t,\theta), \label{ap}\\
j(t,\theta) 
&:=& 
\frac{\pi}{t^2} \int_{-\infty}^\infty \int_0^\infty
w f(t,\theta,w,L)\,dL\,dw =
- e^{\lambda + \mu} T_{01} (t,\theta), \label{aj}\\
q(t,\theta) 
&:=& 
\frac{\pi}{t^4} \int_{-\infty}^\infty \int_0^\infty
\frac{L}{\langle p \rangle} f(t,\theta,w,L)\,dL\,dw =
\frac{2}{t^2}T_{22} (t,\theta);\ \  \label{aq}
\end{eqnarray}
recall that
\[
\langle p \rangle = \sqrt{1+w^2 + L/t^2}.
\]
We prescribe initial data at
some time $t=t_0 >0$, 
\[
f(t_0,\theta,w,L) = \fn (\theta,w,L),\ \lambda (t_0,\theta) = \li (\theta),\ 
\mu (t_0,\theta) = \mi (\theta) 
\]
and want to show that the corresponding solution exists for all $t\geq t_0$.
To this end we make use of the continuation criterion in the
following local existence result:

\begin{prop} \label{locex}
Let $\fn \in C^1(\R^2 \times \R_0^+)$ with 
$\fn (\theta+1 ,w,L)= \fn (\theta,w,L)$ for 
$(\theta,w,L) \in \R^2 \times \R_0^+$, $\fn \geq 0$,
and 
\begin{eqnarray*}
&& \sup \left\{|w| \mid (\theta,w,L) \in \supp \fn \right \} < \infty,\\
&& \sup \left\{ L \mid (\theta,w,L) \in \supp \fn \right \} < \infty .
\end{eqnarray*}
Let $\li,\mi \in C^1 (\R)$
with $\li (\theta) = \li (\theta+1),\ \mi(\theta+1) = \mi(\theta)$ 
for $\theta\in \R$ and
\[
\mi_\theta(\theta) = - 4 \pi e^{\lis + \mis} \open{\jmath}(\theta),\ 
\theta \in \R.
\]
Then there exists a unique, right maximal, regular solution $(f,\lambda,\mu)$
of $(\ref{av})$--$(\ref{af4})$ with $(f,\lambda,\mu)(t_0) = (\fn,\li,\mi)$
on a time interval $[t_0,T[$ with $T \in ]t_0,\infty]$.
If 
\[
\sup \Bigl\{ \mu(t,\theta) \mid \theta \in \R,\ t \in [t_0,T[ \Bigr\} < \infty 
\]
then $T=\infty$.
\end{prop}
This is the content of \cite[Thms.\ 3.1 and 6.2]{rein94a}. In fact this was
only stated in the case $t_0=1$ in that reference, but that was an arbitrary
choice which makes no essential difference. For a {\em regular solution} 
all derivatives which appear in the system exist and are
continuous by definition, cf.\ \cite{rein94a}.

We now establish a series of estimates which will result in an upper bound on
$\mu$ and will therefore prove that $T=\infty$.
Similar estimates were used in \cite{andreasson99a} for the Einstein-Vlasov
system with Gowdy symmetry. In what follows constants denoted
by $C$ will be positive, may depend on the initial data and may change
their value from line to line.

Firstly, integration of (\ref{af2}) with respect to $t$ 
and the fact that $p$ is non-negative imply that
\begin{equation} \label{loweste2mu}
e^{2 \mu(t,\theta)} 
=
\left[ \frac{t_0(e^{-2 \mi(\theta)}+ \epsilon)}{t} - \epsilon - 
\frac{8 \pi}{t} \int_{t_0}^t s^2 p(s,\theta)\, ds\right]^{-1}
\geq 
\frac{t}{C - \epsilon t},\ t \in [t_0,T[.
\end{equation}
Next we claim that
\begin{equation} \label{decrhoint}
\int_0^1 e^{\mu + \lambda} \rho(t,\theta)\, d\theta \leq 
C t^{-1+\epsilon},\ t \in [t_0,T[.
\end{equation}
A lengthy computation shows that
\[
\frac{d}{dt}
\int_0^1 e^{\mu + \lambda} \rho(t,\theta)\, d\theta
=
- \frac{1}{t} \int_0^1 e^{\mu + \lambda}
\biggl[ 2 \rho + q - \frac{\rho + p}{2}(1+\epsilon e^{2 \mu})\biggr]\, 
d\theta .
\] 
Now $q\geq 0$ and $p \leq \rho$ so that for $\epsilon =0$,
\[
\frac{d}{dt} \int_0^1 e^{\mu + \lambda} \rho(t,\theta)\, d\theta 
\leq 
- \frac{1}{t} \int_0^1 e^{\mu + \lambda} \rho(t,\theta)\, d\theta
\]
and integrating this with respect to $t$ yields (\ref{decrhoint}) 
for $\epsilon =0$. For $\epsilon = -1$ we have
\begin{eqnarray*}
\frac{d}{dt} \int_0^1 e^{\mu + \lambda} \rho(t,\theta)\, d\theta 
&\leq&
- \frac{1}{t} \int_0^1 e^{\mu + \lambda}
\biggl[ 2 \rho + q - \frac{\rho + p}{2}\biggr]\, d\theta
- \frac{1}{C+t} \int_0^1 e^{\mu + \lambda} \frac{\rho + p}{2}\, d\theta\\
&\leq&
- \frac{1}{C+t} \int_0^1 e^{\mu + \lambda} (2 \rho + q)\, d\theta
\leq
- \frac{2}{C+t} \int_0^1 e^{\mu + \lambda} \rho\, d\theta,
\end{eqnarray*}
and integrating this inequality with respect to $t$
yields (\ref{decrhoint}) for $\epsilon = -1$.
Using (\ref{decrhoint}) and (\ref{af3}) we find
\begin{eqnarray} \label{muavmu}
\left|\mu (t,\theta) - \int_0^1 \mu(t,\sigma)\, d\sigma \right|
&=&
\left| \int_0^1\int_\sigma^\theta \mu_\theta(t,\tau)\, d\tau\,d\sigma
\right|
\leq
\int_0^1\int_0^1 |\mu_\theta(t,\tau)|\, d\tau\,d\sigma \nonumber \\
&\leq&
4 \pi t \int_0^1 e^{\mu + \lambda}|j(t,\tau)| \, d\tau 
\leq
4 \pi t \int_0^1 e^{\mu + \lambda} \rho(t,\tau) \, d\tau \nonumber\\
&\leq&
C t^{\epsilon},\ t \in [t_0,T[,\ \theta \in [0,1].
\end{eqnarray}
Next we show that
\begin{equation} \label{emuminla}
e^{\mu(t,\theta) - \lambda(t,\theta)} \leq C t^{1+\epsilon},\ t \in [t_0,T[,\ 
\theta \in [0,1].
\end{equation}
To see this observe that by (\ref{af1}), (\ref{af2}) and (\ref{loweste2mu})
\begin{eqnarray*}
\frac{\partial}{\partial t} e^{\mu - \lambda}
&=&
e^{\mu - \lambda}
\left[ 4 \pi t e^{2 \mu} (p-\rho) + (\epsilon e^{2 \mu}+1)/t\right]
\leq
e^{\mu - \lambda} (\epsilon e^{2 \mu}+1)/t\\
&\leq&
\left[\frac{1}{t} + \frac{\epsilon}{C-\epsilon t}\right]\,e^{\mu - \lambda},
\end{eqnarray*}
and integrating this inequality with respect to $t$
yields (\ref{emuminla}).

We now estimate the average of $\mu$ over the interval $[0,1]$
which in combination with (\ref{muavmu}) will yield
the desired upper bound on $\mu$:
\begin{eqnarray*}
\int_0^1 \mu(t,\theta)\, d\theta
&=&
\int_0^1 \mu(t_0,\theta)\, d\theta +
\int_{t_0}^t\int_0^1 \mu_t(s,\theta)\, d\theta\, ds\\
&\leq&
C + \int_{t_0}^t\frac{1}{2 s}
\int_0^1 \left[e^{2 \mu} (8 \pi s^2 p + \epsilon) +1\right]\, d\theta\, ds\\
&=&
C + \frac{1}{2}\ln (t/t_0) 
+ 4 \pi \int_{t_0}^t s
\int_0^1 e^{\mu-\lambda} e^{\mu+\lambda} p \, d\theta\, ds
+ \epsilon \int_{t_0}^t \frac{1}{2 s}
\int_0^1 e^{2 \mu}\, d\theta\, ds\\
&\leq&
C + \frac{1}{2}\ln (t/t_0) + C \int_{t_0}^t s^{2+\epsilon} s^{\epsilon -1} ds
+ \frac{\epsilon}{2} \int_{t_0}^t \frac{ds}{C-\epsilon s},
\end{eqnarray*}
where we used (\ref{loweste2mu}), (\ref{emuminla}) and (\ref{decrhoint}).
With (\ref{muavmu}) this implies
\begin{equation} \label{muest}
\mu(t,\theta) \leq C \left\{
\begin{array}{ccl}
t^2&,& t \in [t_0,T[,\ \theta \in [0,1],\ \epsilon =0,\\
1+\ln t&,& t \in [t_0,T[,\ \theta \in [0,1],\ \epsilon =-1,
\end{array}
\right.
\end{equation}
which by Proposition~\ref{locex} implies $T=\infty$. Thus we have proven:

\begin{theorem} \label{expand}
For initial data as in Proposition~\ref{locex} the corresponding
solution exists for all $t\in[t_0,\infty[$ where $t$ denotes the area
radius of the surfaces of symmetry of the induced spacetime.
The solution satisfies the estimates $(\ref{decrhoint})$, 
$(\ref{emuminla})$ and $(\ref{muest})$.
\end{theorem}

\section{The contracting direction}\label{contract}
\setcounter{equation}{0}

In this section we consider the Einstein-Vlasov system in the contracting
direction and use conformal coordinates in which the metric takes the
form (\ref{cmetric}). As in the case of areal coordinates used in the
expanding direction $f$ depends only on
\[
t,\ \theta,\ p^1,\ (p^2)^2 + \sinh^2 x\, (p^3)^2.
\]
We will write
\[
f=f(t,\theta,w,L)
\]
where
\[
w := e^\eta p^1,\ L:= R^4 ((p^2)^2 + \sinh^2 x\, (p^3)^2).
\]
As above, 
\[
p^0 = e^{-\eta} \sqrt{1+w^2 + L/R^2} =: e^{-\eta} \langle p \rangle, 
\]
and $L$ is a conserved quantity along particle orbits.

In the variables introduced above the Einstein-Vlasov system can be written
in the following form:
\begin{equation} \label{cv}
\partial_t f + \frac{w}{\langle p \rangle}\, \partial_\theta f 
+ \biggl[ - \eta_t w - \eta_\theta\biggl(\langle p \rangle + 
\frac{w^2}{\langle p \rangle}\biggr) + 
e^{-2 \eta} R_\theta \frac{L}{R^3 \langle p \rangle} \biggr]\, \partial_w f 
= 0 .
\end{equation}
\begin{eqnarray}
- R_{\theta\theta} + \eta_t\,R_t +\,\eta_\theta \,R_\theta
+ \frac{1}{2 R}
\Bigl[{R_t}^2 - {R_\theta}^2 - e^{2\,\eta}\Bigr] 
&=&
4 \pi R e^{2\eta} \rho, \label{cf1}\\
R_{t \theta} - \eta_t \, R_\theta - \eta_\theta\, R_t
&=&
4 \pi R e^{2\eta} j, \label{cf2}\\
R_{tt} - R_{\theta \theta} 
+ \frac{1}{R}
\Bigl[{R_t}^2 - {R_\theta}^2 - e^{2\,\eta}\Bigr]
&=&
4 \pi R e^{2\eta}(\rho - p), \label{cf3}\\
\eta_{tt}-\eta_{\theta\theta} - \frac{1}{R^2}
\Bigl[{R_t}^2 - {R_\theta}^2 - e^{2\,\eta}\Bigr]
&=&
4 \pi e^{2\eta} (p - \rho - q), \label{cf4}
\end{eqnarray}
where 
\begin{eqnarray}
\rho(t,\theta)
&:=&
\frac{\pi}{R^2}
\int_{-\infty}^\infty \int_0^\infty 
\langle p \rangle \,f(t,\theta,w,L)\, dL\, dw
= e^{-2\eta} T_{00}(t,\theta), \label{cr}\\
j(t,\theta)
&:=&
\frac{\pi}{R^2}
\int_{-\infty}^\infty \int_0^\infty w f(t,\theta,w,L)\, dL\, dw
= - e^{-2\eta} T_{01}(t,\theta), \label{cj}\\
p(t,\theta)
&:=&
\frac{\pi}{R^2}
\int_{-\infty}^\infty \int_0^\infty \frac{w^2}{\langle p \rangle}
 f(t,\theta,w,L)\, dL\, dw = e^{-2\eta} T_{11}(t,\theta), \label{cp}\\
q(t,\theta)
&:=&
\frac{\pi}{R^4}
\int_{-\infty}^\infty \int_0^\infty \frac{L}{\langle p \rangle} 
f(t,\theta,w,L)\, dL\, dw = \frac{2}{R^2} T_{22}(t,\theta).\label{cq}
\end{eqnarray}
The equations (\ref{cf1}), (\ref{cf2}) are the constraints and
(\ref{cf3}), (\ref{cf4}) are the evolution equations; from the relation
of their right hand sides with the energy momentum tensor it is obvious
how this form of the field equations is obtained.

Let a smooth solution of the system (\ref{cv})--(\ref{cq}) on some
time interval $]t_-,t_0]$ be given. We want to show that
if this interval is bounded and if $R$ is bounded away from zero on
this interval then $f,\ R,\ \eta$ and all their derivatives are
bounded as well, with bounds depending on the data at $t=t_0$
and the lower bound on $R$. To this end we define auxiliary variables
\[
\tau :=\frac{1}{\sqrt{2}}(t-\theta),\ \xi :=\frac{1}{\sqrt{2}}(t+\theta)
\]
so that
\[
\partial_\tau = \frac{1}{\sqrt{2}}(\partial_t-\partial_\theta),\
\partial_\xi = \frac{1}{\sqrt{2}}(\partial_t+\partial_\theta).
\]
The analysis which follows is modeled on the one in \cite{andreasson99a},
cf.\ also \cite{berger97}.

\noindent
{\em Step 1: $C^1$-bounds on $R$}.
A short computation using (\ref{cf1}) and (\ref{cf2}) shows that
\begin{eqnarray*}
\partial_\theta R_\xi
&=&
\frac{1}{\sqrt{2}} (R_{t\theta} + R_{\theta \theta})\\
&=&
\Bigl( \sqrt{2} \eta_\xi + \frac{1}{\sqrt{2} R} R_\tau\Bigr) R_\xi
- \frac{1}{2 \sqrt{2} R} e^{2\eta}  - 2 \sqrt{2} \pi R e^{2\eta} (\rho - j)\\
&<&
\Bigl( \sqrt{2} \eta_\xi + \frac{1}{\sqrt{2} R} R_\tau\Bigr) R_\xi;
\end{eqnarray*}
observe that by (\ref{cr}) and (\ref{cj}), $|j| \leq \rho$. 
Assume that $R_\xi (t,\theta) = 0$ for some $t \in ]t_-,t_0]$ 
and $\theta \in \R$. Then by the periodicity of $R$ with respect to $\theta$,
\[
0=R_\xi(t,\theta +1) < R_\xi(t,\theta) 
\exp\left(\int_\theta^{\theta +1} 
\Bigl( \sqrt{2} \eta_\xi + \frac{1}{\sqrt{2} R} R_\tau\Bigr)\, d\theta'\right)
=0,
\]
a contradiction. Thus $R_\xi \neq 0$ on $]t_-,t_0] \times S^1$.
Similarly,
\[
\partial_\theta R_\tau > 
- \Bigl( \sqrt{2} \eta_\tau + \frac{1}{\sqrt{2} R} R_\xi\Bigr) R_\tau
\]
which yields the same assertion for $R_\tau$. This implies that
the quantity
\[
g^{\alpha \beta} \partial_{x^\alpha} R\, \partial_{x^\beta} R
= e^{-2\eta}(-{R_t}^2 + {R_\theta}^2) = - \frac{1}{2} e^{-2\eta} R_\xi R_\tau
\]
does not change sign. Since $R$ is periodic in $\theta$, there must
exist points where $R_\theta =0$, hence the quantity above is negative
everywhere, and by our choice of time direction,
\begin{equation} \label{gradrtl}
R_t > 0,\ |R_\theta| < R_t\ \mbox{on}\ ]t_-,t_0]\times S^1.
\end{equation}
By (\ref{cf3}),
\[
\frac{1}{2} R_{\xi \tau} = - \frac{1}{2 R} R_\xi R_\tau
+ e^{2\eta} \left(\frac{1}{R} + 4 \pi R (\rho - p)\right) > 
- \frac{1}{2 R} R_\xi R_\tau;
\]
observe that by (\ref{cr}) and (\ref{cp}) $p \leq \rho$.
Now we fix some $(t,\theta) \in ]t_-,t_0[\times \R$. Then
for $s \in [t,t_0]$,
\[
\frac{d}{ds} R_\xi (s,\theta +t-s)
=
\sqrt{2} R_{\xi \tau} 
>
- \frac{d}{ds} [\ln R(s,\theta +t-s)] \,R_\xi (s,\theta +t-s).
\]
Integrating this differential inequality yields
\[
R_\xi (t,\theta) < 
\frac{R(t_0,\theta+t-t_0)}{R(t,\theta)} R_\xi (t_0,\theta +t - t_0).
\]
Similarly,
\[
R_\tau (t,\theta) < 
\frac{R(t_0,\theta-t+t_0)}{R(t,\theta)} R_\tau (t_0,\theta -t + t_0),
\]
and both estimates together imply that $R_t$ is bounded from above
on $]t_-,t_0]\times S^1$ with a bound of the desired sort.
Together with (\ref{gradrtl}) this provides bounds for
$R_t$ and $R_\theta$. Note that this argument shows that the spacetime 
gradient of $R^2$ is bounded even without the assumption that $R$ is 
bounded away from zero.

\noindent
{\em Step 2: $C^1$-bounds on $\eta$}.
From (\ref{cf3}) and (\ref{cf4}) we find
\[
\eta_{\xi \tau} = - \frac{1}{2 R} R_{\xi\tau}
- 2 \pi e^{2\eta} q.
\]
We fix some $(t,\theta) \in ]t_-,t_0[\times \R$. Then
for $s\in[t,t_0]$,
\[
\frac{d}{ds} \eta_\xi (s,\theta +t-s)
=
- \frac{1}{R} \frac{d}{ds} R_\xi (s,\theta +t-s)
- 2 \sqrt{2} \pi e^{2\eta} q(s,\theta +t-s).
\]
Integrating this and integrating the term containing $R_\xi$ by
parts yields
\begin{eqnarray} \label{etaxirep}
\eta_\xi (t,\theta) 
&=& 
\eta_\xi (t_0,\theta + t - t_0)
- \frac{R_\xi (t_0,\theta + t - t_0)}{R (t_0,\theta + t - t_0)} 
 + \frac{R_\xi (t,\theta)}{R (t,\theta)} \nonumber \\
&& 
{} + 
\sqrt{2} \int_t^{t_0} \left(2 \pi e^{2\eta} q -
 \frac{R_\tau R_\xi}{R^2}\right)(s,\theta +t-s)\, ds.
\end{eqnarray}
We already know that
\[
\int_t^{t_0} R_{\xi\tau}(s,\theta +t-s)\, ds =
\frac{1}{\sqrt{2}} \left(R_\xi(t_0,\theta + t - t_0) - R_\xi(t,\theta)\right)
\]
is bounded, and on the other hand by (\ref{cf3})
the left hand side can be written as
\[
2 \int_t^{t_0}\left[\frac{1}{R} \Bigl(-{R_t}^2 + {R_\theta}^2\Bigr)
+ \frac{1}{R} e^{2\eta}
+ 4 \pi R e^{2\eta}(\rho - p)\right](s,\theta +t-s)\, ds.
\]
Here the first term is bounded by Step 1, and the second and third terms
are non-negative. Thus by (\ref{cq}), (\ref{cr}), (\ref{cp}),
\[
\int_t^{t_0} e^{2\eta} q(s,\theta +t-s)\, ds
\leq
\int_t^{t_0}\frac{1}{R^2}e^{2\eta} (\rho - p)(s,\theta +t-s)\, ds
\]
is bounded as well; recall that $R$ is 
bounded away from zero by assumption. Thus (\ref{etaxirep})
implies that $\eta_\xi$ is bounded on $]t_-,t_0]\times S^1$.
Analogously to (\ref{etaxirep}) we have
\begin{eqnarray} \label{etataurep}
\eta_\tau (t,\theta) 
&=& 
\eta_\tau (t_0,\theta - t + t_0)
- \frac{R_\tau (t_0,\theta - t + t_0)}{R (t_0,\theta - t + t_0)}  + 
\frac{R_\tau (t,\theta)}{R (t,\theta)} \nonumber \\ 
&&
{} + 
\sqrt{2} \int_t^{t_0} \left( 2 \pi e^{2\eta} q -
\frac{R_\tau R_\xi}{R^2}\right) (s,\theta -t+s)\, ds
\end{eqnarray}
from which we can conclude that $\eta_\tau$
is bounded on $]t_-,t_0]\times S^1$ as well. Thus also
$\eta_t, \eta_\theta$ and $\eta$ itself are bounded there.

\noindent
{\em Step 3: Bounds on matter quantities}.
We have for any $t \in ]t_-,t_0],\ \theta \in \R,\
w \in \R,\ L> 0$,
\[
f(t,\theta,w,L) = f(t_0,\Theta(t_0,t,\theta,w,L),W(t_0,t,\theta,w,L),L)
\]
where $\Theta(\cdot,t,\theta,w,L),W(\cdot,t,\theta,w,L)$
is the solution of the characteristic system 
\begin{eqnarray*}
\dot \theta 
&=&
\frac{w}{\langle p \rangle},\\
\dot w 
&=&
- \eta_t w - \eta_\theta\biggl(\langle p \rangle + 
\frac{w^2}{\langle p \rangle}\biggr) + 
e^{-2 \eta} R_\theta \frac{L}{R^3 \langle p \rangle}
\end{eqnarray*}
of the Vlasov equation with 
$\Theta(t,t,\theta,w,L)= \theta,\ W(t,t,\theta,w,L)= w$.
This representation of $f$ implies immediately that
$f$ remains non-negative and bounded by its maximum at $t=t_0$.
By Steps 1 and 2 the right hand side of the second
equation in the characteristic system is linearly bounded
in $w$, and hence if the $w$-support of $f$ is
compact initially, 
\[
\sup\{ |w| | (\theta,w,L) \in \supp f(t),\ t \in ]t_-,t_0]\} < \infty .
\] 
This immediately implies that $\rho, p, j, q$
are bounded on $]t_-,t_0] \times S^1$.

\noindent
{\em Step 4: Bounds on second order derivatives of $R$ and $\eta$}.
Steps 1, 2 and 3 together with (\ref{cf1}), (\ref{cf2}) and (\ref{cf3})
imply than $R_{\theta \theta}, R_{t \theta}$ and $R_{tt}$ are bounded as
claimed on $]t_-,t_0]\times S^1$. The bounds on the second order
derivatives of $\eta$ are a bit less trivial: We add the equations 
(\ref{etaxirep}) and (\ref{etataurep}) to obtain a formula for 
$\eta_\theta$. When this formula is differentiated with respect to $\theta$
there results a number of terms which are bounded by the previous steps
and the terms
\begin{eqnarray}
&&
2 \pi \int_t^{t_0} e^{2\eta} \frac{\pi}{R^4}
\int_{-\infty}^\infty\int_0^\infty \frac{L}{\langle p\rangle}
\partial_\theta f(s,\theta+t-s,w,L)\,dL\,dw\,ds, \label{ftheta1}\\
&&
2 \pi \int_t^{t_0} e^{2\eta} \frac{\pi}{R^4}
\int_{-\infty}^\infty \int_0^\infty \frac{L}{\langle p\rangle}
\partial_\theta f(s,\theta-t+s,w,L)\,dL\,dw\,ds \label{ftheta2}.
\end{eqnarray}
To deal with the first term we introduce the differential operators
\[
W=\sqrt{2}\, \partial_\tau = \partial_t - \partial_\theta,\
S=\partial_t + \frac{w}{\langle p\rangle} \partial_\theta
\]
so that
\[
\partial_\theta = \frac{\langle p\rangle}{\langle p\rangle +w} (S-W).
\]
By the Vlasov equation
\[
Sf = - \left[- \eta_t w - \eta_\theta\biggl(\langle p \rangle + 
\frac{w^2}{\langle p \rangle}\biggr) + 
e^{-2 \eta} R_\theta \frac{L}{R^3 \langle p \rangle}\right]\, \partial_w f.
\]
When this is substituted into (\ref{ftheta1}) the resulting 
term can be integrated
by parts with respect to $w$, and all the terms which then appear are bounded
by the previous steps. As to the $W$-contribution,
\[
(Wf)(s,\theta+t-s,w,L) = \frac{d}{ds}\left[f(s,\theta +t-s,w,L)\right]
\]
so that the corresponding term in (\ref{ftheta1}) can be integrated by
parts with respect to $s$ which again results in bounded terms. 
In order to deal
with (\ref{ftheta2}) we replace $w$ by $-w$ in the integral and redefine
\[
W = \partial_t + \partial_\theta,\
S = \partial_t - \frac{w}{\langle p\rangle} \partial_\theta;
\]
the rest of the argument should then be obvious, and $\eta_{\theta\theta}$
is seen to be bounded on $]t_-,t_0]\times S^1$. 
This way of dealing with derivatives
in connection with the Vlasov equation was introduced for the Vlasov-Maxwell
system in \cite{GS}. Now that $\eta_{\theta\theta}$ is bounded
the same is true for $\eta_{tt}$ by (\ref{cf4}), and $\eta_{t\theta}$
can be dealt with like $\eta_{\theta\theta}$: Taking the difference
of (\ref{etaxirep}) and (\ref{etataurep}) gives the necessary
formula for $\eta_t$.

\noindent
{\em Step 5: Higher order derivatives.}
Via the characteristic system the $C^2$-bounds on $R$ and $\eta$ 
give bounds on the first order
derivatives of $\Theta(\cdot,t,\theta,w,L)$ and $W(\cdot,t,\theta,w,L)$
with respect to $\theta,w,L$. This yields corresponding
$C^1$-bounds first on $f$ and then as in Step 3 on $\rho, p, j, q$.
These in turn imply $C^3$-bounds on $R$. The third order derivatives of $\eta$
then have to be dealt with by repeating the argument of Step 4,
and although the details would be tedious it should be clear that this
process can be iterated to bound any desired derivative on
$]t_-,t_0]\times S^1$ in terms of the data at $t=t_0$ and the positive lower
bound on $R$.

Later we will require a slight generalization of these results. The essence
of the first part of Step 1 above is to show that the gradient of $R$ is 
timelike. The argument is carried out on a region which is covered by
Cauchy surfaces of constant conformal time. However this fact is also true for
any solution of the Einstein-Vlasov system with hyperbolic symmetry and a 
compact Cauchy surface. This follows from \cite[Lemma~2.5]{rendall95a}.  
Once this has been established the estimates in the later steps hold for 
any subset $Z$ of the half-plane $t\le t_0$ provided $Z$ is a future set. 
By definition this means that any future directed causal curve in the 
region $t\le t_0$ starting at a point of $Z$
remains in $Z$. (For information on concepts such as this 
concerning causal structures see e.g. \cite{hawking73}.) Thus if $R$ is 
bounded away from zero on $Z$ and $t$ is bounded on $Z$ then all the 
unknowns and their derivatives can be controlled on $Z$.

Now consider a special choice of the subset $Z$, namely that which is 
defined by the inequalities $t_1<t\le t_0$ and 
$\theta_1+t_0-t<\theta<\theta_2-t_0+t$ for some numbers $\theta_1$, 
$\theta_2$ and 
$t_1$ satisfying the inequalities $\theta_1<\theta_2$ and $t_1>t_0-(1/2)
(\theta_2-\theta_1)$.
Suppose a solution of the Einstein-Vlasov system in conformal coordinates
defined on this region is such that $R$ is bounded away from zero. Then the
functions defining the solution extend smoothly to the boundary of $Z$ at
$t=t_1$. They define smooth Cauchy data for the Einstein-Vlasov system.
Applying the standard local existence theorem (without symmetry) allows 
the solution to be extended through that boundary. Repeating the 
construction of conformal coordinates in Section~\ref{hyperbolic} then
shows that we get an extension of the solution written in conformal 
coordinates through that boundary.

\section{Existence of an areal time coordinate}\label{cauchy}

In Section~\ref{hyperbolic} it was shown that the maximal Cauchy development
of initial data for the Einstein-Vlasov system with hyperbolic symmetry
admits conformal coordinates in a neighbourhood of the initial 
hypersurface. The aim of this section is to show that a spacetime of this
kind always admits a symmetric compact Cauchy surface of constant areal time.
The basic strategy follows that of \cite{berger97}.  
The fundamental object of study is a spacetime in conformal coordinates
which develops (to the past) from initial data on the hypersurface $t=0$.
The solution will exist on a subset of the region $t\le 0$ of $\R^2$.
We will consider solutions defined on open subsets of the half-plane 
$t\le 0$ containing $t=0$ which are future sets.  The union of all regions
admitting solutions of this type is an open future set. Denote it by 
$\cal D$. By definition $R$ is a positive function defined everywhere on
$\cal D$.

Consider first the case that the past boundary of $\cal D$ is empty, i.e., 
that 
${\cal D}=]-\infty,0]\times S^1$. Let $\Sigma_\rho$ be the level set defined
by the equation $R(t,\theta)=\rho$ for some $\rho$ less than the minimum of
$R$ on $t=0$. Since the gradient of $R$ is everywhere timelike the level 
sets $\Sigma_\rho$ are smooth spacelike submanifolds. These level sets
can be represented in the form $t=f(\theta)$, and since the coordinates are 
conformal and the level sets spacelike it follows that $|f'|<1$. Hence
the $\Sigma_\rho$ are compact and their pull-backs define compact spacelike 
hypersurfaces in a Cauchy development of the initial data of interest.
By \cite{budic78} these are Cauchy surfaces. As a consequence the maximal 
Cauchy development of those data contains symmetric compact Cauchy surfaces
of constant areal time.

It remains to consider the case where $\cal D$ has a non-empty past
boundary. Call this boundary $\cal B$. It is achronal. Hence for any
given $\theta$ there is only one $t$ such that $(t,\theta)\in\cal B$. If $p$ 
is a 
point of $\cal B$ then all points of $\cal B$ close to $p$ are outside 
the light cone of $p$. Hence $\cal B$ is a Lipschitz curve with Lipschitz 
constant one. In other words it can be represented in the form $t=h(\theta)$
for a function $h$ with Lipschitz constant one. This is a special case
of a general property of achronal boundaries, cf.\
\cite[p.~187]{hawking73}.
As a consequence $\cal B$ is compact. Since the gradient of $R^2$
is bounded on the whole region where the solution is defined (as shown
in Section~\ref{contract}) it follows that $R^2$ is uniformly continuous.
Hence the function $R^2$ extends to a continuous function 
on the closure $\bar{\cal D}$ of $\cal D$. Its square root $\bar R$ is
a continuous extension of $R$ to $\bar{\cal D}$. It will now 
be proved that $\bar R=0$ on $\cal B$. From this point on it is possible 
to argue as above to obtain the existence of a Cauchy surface on which 
$R$ is constant.

Consider first a point $\theta_0$ where $h$ has a local maximum and let 
$t_0=h(\theta_0)$. Then there exists $\epsilon>0$ such that for any $t$ greater
than $t_0$ the points $(t,\theta)$ with 
$\theta_0-\epsilon<\theta<\theta_0+\epsilon$ 
lie in $\cal D$. If $\bar R(t_0,\theta_0)>0$ then the solution has a smooth 
limit at the points $(t_0,\theta)$ with 
$\theta_0-\epsilon<\theta<\theta_0+\epsilon$ provided $\epsilon$
is small enough. This gives Cauchy data at $t=t_0$. The corresponding local 
solution of the evolution equations can be used to extend the original 
solution, thus contradicting the assumed maximality of $\cal D$. It follows
that in fact $\bar R$ vanishes at any point of $\cal B$ where $h$ has a local 
maximum.

Let ${\cal D}_t$ be the intersection of $\cal D$ with the hypersurface where
the time coordinate takes the value $t$. It is an open subset of $S^1$. It
is either the whole of $S^1$, the empty set or a disjoint union of 
open intervals which are its connected components. Consider now a point 
$(t_0,\theta_0)$ of $\cal B$ which is an endpoint of a component of 
${\cal D}_{t_0}$. Suppose further that $\bar R(t_0,\theta_0)>0$. 
Without loss of generality we can assume 
that $\theta_0$ is a right endpoint of a component; the argument for 
a left endpoint is strictly analogous. For some $\epsilon>0$ the solution 
extends smoothly to the points $(t_0,\theta)$ with 
$\theta_0-\epsilon<\theta\le \theta_0$. This 
can be extended to smooth data at $t=t_0$ on the interval 
$\theta_0-\epsilon<\theta< \theta_0+\epsilon$. 
There is a corresponding local solution. 
Using the domain of dependence we see that this new solution extends the 
original solution at those points outside the past light cone of the point 
$(t_0,\theta_0)$. In this way we obtain an extension of the original solution 
and a contradiction unless the part of $\cal B$ immediately to the left of 
$\theta_0$ is a null curve. If ${\cal B}_1$ is the set of points of $\cal B$ 
which are endpoints of components of some ${\cal D}_t$ and which have 
$\bar R >0$ then it has now been shown that ${\cal B}_1$ is a union of null 
segments. Returning to the point $(t_0,\theta_0)$ we can ask what happens 
to the interval of which it is an endpoint as $t$ increases. The endpoint 
could vanish if the interval coalesces with another while $\bar R$ remains 
positive. However that would mean that $\theta_0$ would 
have to be a local maximum of $h$, a case already excluded. Hence as we move 
to the right in $\theta$ the corresponding point of $\cal B$ must continue 
to be an endpoint. Eventually $\bar R$ must become zero. But in that case the 
future-pointing character of the gradient of $R$ can be used  to obtain a 
contradiction. It can be concluded that ${\cal B}_1$ is empty.

It remains to consider the case that there is a point $(t_0,\theta_0)$ of 
$\cal B$ which is not an endpoint of a component of some ${\cal D}_t$. In
this case $(t,\theta_0)$ belongs to $\cal D$ for every $t>t_0$. Let 
$(a(t),b(t))$ be the component of ${\cal D}_t$ containing $\theta_0$. The
functions $a(t)$ and $b(t)$ are decreasing and increasing, respectively.
Let $a_0$ and $b_0$ be their limits as $t\to t_0$ from above. If $a_0<b_0$
then $(t_0,\theta_0)$ corresponds to a local maximum of $f$ and so $\bar R=0$ 
there. Otherwise it is a limit of the points $(t,b(t))$ which are right 
endpoints of components of ${\cal D}_t$, and so by continuity $\bar R=0$. 

A natural question arising in this context is whether a spacetime which
exists globally in one time direction with respect to a conformal time
coordinate cannot be extended in that direction to a strictly larger
globally hyperbolic spacetime. It will be shown that this is impossible.
As an aid in doing this we consider the action of the symmetry group on
$\tilde M$. We know exactly what the action looks like near the initial 
hypersurface. There is a two-dimensional subgroup $H$ of $G$ which acts 
transitively on the orbits of the action of $G$. Corresponding to $H$
there are two Killing vector fields which are spacelike and linearly
independent near the initial hypersurface. We claim that they are
spacelike and linearly independent on the whole of $\tilde M$, so that 
the orbits of the action of $H$ are everywhere spacelike and
two-dimensional. To show this it suffices to show that a Killing
vector generated by the action of $H$ can neither vanish nor become null
anywhere on $\tilde M$, cf.\ \cite[Prop.~2.4]{andersson99a}. 
Suppose that a Killing
vector of this kind vanishes at a point $p$. 
Without loss of generality we can assume 
that $p$ lies to the future of the initial hypersurface and that the
Killing vector is spacelike and non-zero on the chronological past
of $p$. The integral curve of the Killing vector through $p$ is null
and if we extend it maximally it must either reach the initial
hypersurface or tend to a zero of the Killing vector. The first of 
these possibilities contradicts the fact that the Killing vector is 
spacelike on the initial hypersurface. In the second case we see
that the Killing vector has a zero to the future of the initial surface.

Now it will be shown that the existence of zero of the Killing vector
at a point $p$ to the future of the initial hypersurface leads to a 
contradiction. The flow generated by the Killing vector leaves
$p$ invariant. Its linearization maps the space of null vectors at
$p$ to itself and must have a fixed point. In other words the 
linearization leaves a null direction invariant. The null geodesic  
with this initial direction is then also invariant. It must intersect
the initial hypersurface, and the point of intersection is invariant
under the flow. This gives the desired contradiction.

Now we come back to the question of inextendibility. Suppose that the
spacetime exists for all conformal time in one time direction, 
without loss of generality
the future. Suppose that there is a point $p$ belonging to a future
development of the initial data but not in the region covered by the
conformal coordinates. It lies on an orbit of $H$. Let $\gamma_1$
and $\gamma_2$ be the two null geodesics through $p$ orthogonal to the 
orbit. They are orthogonal to all orbits they meet. They must enter 
the region covered by conformal coordinates and then they coincide
with the curves with $t=\pm \theta+C$ for some constant $C$ and constant 
values of the other coordinates. Without loss of generality
we can assume that $p$ is on the 
boundary of the region covered by the conformal coordinates. In the 
approach to $p$ the geodesic $\gamma_2$ intersects the geodesic $\gamma_1$
infinitely many times. This contradicts the strong causality and
hence the global hyperbolicity of the extension. The conclusion 
is that an interval of conformal time which is infinite in one time
direction proves that the maximal Cauchy development is exhausted
in that time direction.  

There is also another criterion for inextendibility which can conveniently
be discussed here. It is related to an argument given in \cite{berger97}.
Consider the action of the group $H$ on $\tilde M$ introduced above and
the corresponding Killing vectors. Since they are always linearly 
independent with spacelike orbits, the $2\times 2$ matrix of inner products 
of these vectors always has positive determinant $\Delta$. Suppose
that a solution is given which exists globally towards the future in areal
time. It will now be shown that it has no proper globally hyperbolic extension
to the future. If there was such an extension there would be a point $p$
of the extension not belonging to the original spacetime. Let $\gamma_1$ be 
a null geodesic through $p$ as above. We may assume without loss of 
generality that it immediately enters the original spacetime on leaving $p$ 
towards the past. In the region covered by the areal time coordinate it 
has constant values of the coordinates $x$ and $y$. Along the part of 
$\gamma_1$ belonging to the original spacetime $\Delta$ is proportional to 
$R^4$. As a point on $\gamma_1$ approaches $p$ the areal time coordinate 
$t$ tends to infinity, which means that $\Delta$ tends to infinity. But 
$\Delta$ is a smooth function at the point $p$. Hence no point $p$ of this 
kind can exist.  

\section{Main results}\label{results}
\setcounter{equation}{0}

In this section the analytical and geometrical information obtained in
previous sections is combined to obtain the main results of the 
paper.

\begin{theorem}\label{theorem1} 
Let $(M,g_{\alpha\beta},f)$ be the maximal 
globally hyperbolic development of initial 
data for the Einstein-Vlasov system with hyperbolic symmetry. 
Then $M$ can be covered by symmetric compact hypersurfaces of constant 
area radius. The area radius of these hypersurfaces takes all values in 
the range $]R_0,\infty[$ where $R_0$ is a non-negative number. 
\end{theorem}

The quantity $R_0$
depends on the solution, but there is no good understanding of how
many solutions have $R_0\ne 0$.

\noindent{\em Proof of Theorem~\ref{theorem1}.}
For a spacetime satisfying the hypotheses of the theorem we
know from Section~\ref{hyperbolic} that a conformal coordinate system
can be introduced on a neighbourhood of the initial hypersurface $S_0$
corresponding to the original data. By the results of Section~\ref{cauchy} 
it follows that this region can be extended to the past so as to include 
a Cauchy surface $S_A$ of constant areal time. Moreover, either the 
conformal time coordinate extends to all negative values, or $R$ tends to 
zero as the past boundary $\cal B$ of the region covered by conformal
coordinates is approached. In the first of these cases the region
covered by the conformal time coordinate includes the entire past of the
initial hypersurface in the maximal Cauchy development, as shown in
Section~\ref{cauchy}. Also the past of $S_A$ in that region admits a 
foliation by hypersurfaces of constant $R$. In that region we can transform 
to areal coordinates. For we can choose the spatial coordinate $\theta$ so 
that its coordinate lines in $Q$ are orthogonal to that foliation. In the 
second case (where $R$ tends to zero on a boundary $\cal B$) the past of 
$S_A$ is also covered by areal coordinates. It exhausts the past of $S_A$ 
in the maximal Cauchy development, as will now be shown. A well-known
argument related to Hawking's singularity theorem shows that it is enough
to check that the mean curvature of the foliation tends uniformly to 
infinity as $R$ tends to zero. The mean curvature is given in areal 
coordinates by \cite{rein94a}
\begin{equation}
\tr k=-e^{-\mu}(\dot\lambda+2/t)
\end{equation} 
Substituting the field equation for $\dot\lambda$ and using the fact that
$\rho\ge 0$ gives the inequality
\begin{equation}\label{crush}
\tr k\le -(1/2t)(3e^{-\mu}+e^\mu)\le -\sqrt 3/t
\end{equation}
and hence the desired result. Thus in both cases the entire past of $S_0$ 
in the maximal Cauchy development is covered. It follows from the results 
of Section~\ref{expandd} that the spacetime and the areal time coordinate 
can be extended so that the time coordinate covers the interval 
$]R_0,\infty[$. It can then be concluded by the argument at the end of 
Section~\ref{cauchy} that the entire future of $S_0$ in the maximal Cauchy 
development is covered.

\begin{theorem}\label{theorem2}
Let $(M,g_{\alpha \beta},f)$ be the maximal globally hyperbolic 
development of initial data for 
the Einstein-Vlasov system with hyperbolic symmetry. Then $M$ can be 
covered by compact hypersurfaces of constant mean curvature. The mean 
curvature of these hypersurfaces takes all values in the range 
$]-\infty, 0[$.
\end{theorem}

\noindent
{\em Proof.}
It follows from (\ref{crush}) that there are Cauchy surfaces
with everywhere negative mean curvature. Under these circumstances it
was shown by Henkel \cite{henkel01c} that the initial singularity is
a crushing singularity and thus a neighbourhood of it can be foliated by 
CMC hypersurfaces. Next the statement about the range of the CMC time 
coordinate will be proved. The important observation is that
the argument used to prove the corresponding statement in \cite{rendall97a}
extends straightforwardly to the case of hyperbolic symmetry. It is 
more powerful than the arguments used to extend the CMC foliation in
\cite{rendall95a}. It remains to see that the CMC foliation covers the 
entire future of the initial hypersurface. It is enough to show that if 
$p$ is any point of the spacetime there is a compact CMC hypersurface 
which contains $p$ in its past. Let $S_1$ be the Cauchy surface of constant 
areal time passing through $p$. Equation (\ref{crush}) shows that the 
mean curvature of $S_1$ is strictly negative. Hence it has a maximum value
$H_1<0$. Let $S_2$ be the compact CMC hypersurface with mean curvature
$H_1/2$. Then the infimum of the mean curvature of $S_2$ is greater
than the supremum of the mean curvature of $S_1$ and a standard argument
\cite{marsden80} shows that $S_2$ is strictly to the future of $S_1$. Hence 
$p$ is in the past of $S_2$, as required.

\noindent
{\em Remark.}
In the case of Gowdy symmetry the mean curvature of the 
hypersurfaces of constant areal time is negative and the corresponding 
argument applies, thus showing that the CMC foliation covers the spacetime 
in that case too. 

The considerations of this paper were confined to the case
where the matter content of spacetime is described by the Vlasov equation,
but in fact many of the arguments do not depend on the details of the 
matter model. It would be worth to investigate which parts of the 
conclusions extend to which matter models. Some relevant estimates for
scalar fields, wave maps and electromagnetic fields can be found in
\cite{rendall95a}, \cite{burnett}, \cite{rendall97a} and \cite{henkel01c}. 

\section{Possible further developments}

In this paper we have seen how it is possible to get rather complete
information on the existence of global geometrically defined time
coordinates in spacetimes with hyperbolic symmetry and, more generally,
with surface symmetry. The motivation for being interested in this
question was that its answer could provide a tool for understanding
the global geometry of these spacetimes. To do so it is necessary
to understand something about the asymptotic behaviour of the solutions
as expressed with respect to these time coordinates. The kind of 
questions which we would like to answer are the following. Is the
spacetime geodesically complete towards the future? Is there a
curvature singularity in the past? Is the singularity in the past
velocity dominated? Does the solution become homogeneous in the
future in some sense? Can we obtain detailed asymptotic expansions
for the solution in these regimes?

The one strong and rather general result concerning questions like
this which is available is the following theorem of \cite{rein94a}. If 
there is a foliation by Cauchy surfaces of constant $R$ such
that $R$ approaches zero as the singularity is approached and
if, in the hyperbolic case, the Hawking mass is positive near 
the singularity then the Kretschmann scalar blows up as the 
singularity is approached. That some restriction is required is
shown by the pseudo-Schwarzschild solution (cf. \cite{rendall95a}). For 
negative values of the mass parameter this solution, which has hyperbolic
symmetry, has negative Hawking mass, and $R$ does not approach 
zero at the boundary of the maximal Cauchy development. The
Kretschmann scalar remains finite near that boundary. A guess,
which is not contradicted by any known results, is that the
pseudo-Schwarzschild solutions are the only solutions of the 
Einstein-Vlasov system with hyperbolic symmetry which do not
have a singularity where the Kretschmann scalar blows up. It
would be very interesting to have a rigidity theorem of this 
kind.

Apparently the only other source of information about the asymptotics
of surface symmetric solutions of the Einstein-Vlasov system
comes from the study of spatially homogeneous solutions. The
spatially homogeneous solutions with spherical, plane and 
hyperbolic symmetry are the spacetimes of Kantowski-Sachs,
Bianchi I and Bianchi III types respectively. There are exceptional
cases (all vacuum) which are the Schwarzschild solution (appropriately
identified), the flat Kasner solution and the pseudo-Schwarzschild
solution. These were presented in Appendix B to \cite{rendall95a}.
All other solutions are such that curvature invariants blow up
in both time directions (Kantowski-Sachs) or that curvature 
invariants blow up in one time direction and the spacetime is
geodesically complete in the other time direction (Bianchi types
I and III) \cite{rendall95b}. 
 
More detailed results on the asymptotics have been obtained for
Bianchi types I and III \cite{rendall00a}. Translating the theorems
obtained there into the notation of this paper leads to the
following results. All non-vacuum solutions of either of these two Bianchi 
types are such that $R\to 0$ as the past boundary of the maximal 
Cauchy development is approached. Generically $\mu\sim (1/2)\ln t$ and
$\lambda\sim -(1/2)\ln t$ as $t\to 0$, where we only indicate the leading 
term in an asymptotic expansion. There is a smaller class of solutions
(not explicitly known) where $\mu\sim (1/2)\ln t$ and 
$\lambda\sim {\rm const.}$ as $t\to 0$. Finally there is an even smaller 
class where $\mu\sim \ln t$ and $\lambda\sim \ln t$ as $t\to 0$. In the 
expanding direction all solutions of both Bianchi types are geodesically 
complete. For type I we have $\mu\sim (1/2)\ln t$ and $\lambda\sim \ln t$ 
as $t\to\infty$. The behaviour of the type III solutions was not completely 
determined in \cite{rendall00a}. It was shown that (in the notation of the 
present paper) $\lambda$ is increasing for large $t$ but probably it grows 
slower than any positive multiple of $\ln t$.

Clearly the next step is to extend these results on asymptotic
behaviour to the inhomogeneous case. It would be convenient if
the inhomogeneous solutions behaved essentially like the homogeneous
ones in their asymptotic regimes. There are, however, phenomena
which may prevent this. These are the possibility of the 
occurrence of analogues of the spikes found in Gowdy spacetimes 
\cite{rendall01a} in the contracting direction and the Jeans instability 
\cite{boerner93} in the expanding direction.

{\em Acknowledgement:} GR acknowledges support by the Wittgenstein 
2000 Award of P.~A.~Markowich.

\end{document}